\documentstyle[12pt]{article}
\author{Afraimovich~E.~L., Perevalova~N.~P.,Voyeikov~S.~V.}
\title{\large
{\vspace{3ex}
Traveling wave packets of total electron content disturbances
as deduced from global GPS network data}}
\date{}
\begin{document}
\maketitle

\begin{abstract}
We identified a new class of mid-latitude medium-scale
traveling ionospheric disturbances (MS TIDs), viz. traveling wave
packets (TWPs) of total electron content (TEC) disturbances. For
the first time, the morphology of TWPs is presented for 105 days
from the time interval 1998-2001 with a different level of
geomagnetic activity, with the number of stations of the global
GPS network ranging from 10 to 300. The radio paths used in the
analysis total about 700000. These data were obtained using the
GLOBDET technology for global detection and monitoring of
ionospheric disturbances of natural and technogenic origin from
measurements of TEC variations acquired by a global network of
receivers of the navigation GPS system. The GLOBDET technology
was developed at the ISTP SB RAS. Using the technique of GPS
interferometry of TIDs we carried out a detailed analysis of the
spatial-temporal properties of TWPs by considering an example of
the most conspicuous manifestation of TWPs on October 18, 2001
over California, USA. It was found that TWPs are observed no
more than in 0.1-0.4\% of all radio paths, most commonly during
the daytime in winter and autumn at low geomagnetic activity.
TWPs in the time range represent quasi-periodic oscillations
of TEC of a length on the order of 1 hour with the oscillation
period in the range 10-20 min and the amplitude exceeding the
amplitude of "background" TEC fluctuations by one order of
magnitude, as a minimum. The radius of spatial correlation
of TWPs does not exceed 500--600 km (3--5 wavelengths). The velocity
and direction of TWPs correspond to those of mid-latitude
medium-scale traveling ionospheric disturbances (MS TIDs)
obtained previously from analyzing the phase characteristics
of HF radio signals as well as signals from geostationary satellites
and discrete cosmic radio sources.
\end{abstract}

\section{Introduction}
\label{SPE-sect-1}

The unremitting interest in investigations of atmospheric
acoustic-gravity waves (AGW) over more than four decades dating
back to Hines pioneering work (Hines, 1960, 1967) is dictated by
the important role played by these waves in the dynamics of the
Earth's atmosphere. These research efforts have been addressed in
a large number of publications, including a series of through
reviews (Hocke and Schlegel, 1996; Oliver et al., 1997).

AGW typically show up in the ionosphere in the form of traveling
ionospheric disturbances (TIDs) and are detected by various radio
techniques. TIDs are classified as large- and medium-scale
disturbances differing by their horizontal phase velocity which
is larger (in the large-scale case) or smaller (for the medium
scale) velocity of sound in the lower thermosphere (on the order
of 300 m/s), with periods within 0.5--3.0 h and 10--40 min,
respectively. Medium-scale TIDs (MS TIDs) are observed
predominantly during the daytime hours and are associated with
AGW which are generated in the lower atmosphere. Large-scale TIDs
are predominant in the night-time hours and are closely
associated with geomagnetic and auroral activity.

It is known that the sources of medium-scale AGW can include
natural processes of a different origin: magnetic storms, auroral
phenomena, weather factors, tropospheric turbulence and jet
flows, the solar terminator, strong earthquakes, volcanic
eruptions, as well as anthropogenic influences (rocket
launchings, explosions, nuclear tests). As a consequence the
observed picture of the electron density disturbance is
essentially a net interference wave field of the AGW of a
different origin. Identifying of the AGW of a definite type from
this field is a highly involved and generally an almost
intractable problem.

The most reliable measurements of the main parameters of
medium-scale AGW (parameters of the wave vector of the AGW,
spectral and dispersion characteristics, etc.) can therefore be
made only for a very rare, unusual type of MS TIDs, i.e.
quasi-periodic (monochromatic) oscillations which are sometimes
recorded as corresponding variations of the frequency Doppler
shift $F_D$ of the ionosphere-reflected HF radio signal (Davies and
Jones, 1971; Waldock and Jones, 1987; Jacobson and Carlos, 1991;
Yakovets et al., 1999).

Experiments of this kind were instrumental in investigating the
spatial-temporal characteristics of MS TIDs in the form of a wave
process, because such recordings are easy to identify visually
with monochromatic individual AGW. Unfortunately, this was
possible to accomplish for a very limited body of experimental
data. Thus, Jacobson and Carlos (1991) managed to identify only a
few monochromatic TIDs from their data obtained for more than 100
hours of observation.

Yakovets et al. (1999) also recorded only a few realizations of
monochromatic TIDs for two observing periods from the winter
months of 1989 and 1990. Yakovets et al. (1999) are likely to be
the first to use the term "wave packets" to designate the
quasi-monochromatic variations of $F_D$, and they made an attempt to
explain their origin on the basis of studying the phase structure
of the oscillations. The authors of the cited reference observed
two types of $F_D$-variations: quasi-stochastic TIDs, and
monochromatic TIDs in the form of wave packets. They arrived at
the conclusion that quasi-stochastic TIDs are characterized by a
random phase behavior, a short length of coherence, and by a
large vertical phase velocity. Wave packets show
quasi-monochromatic oscillations of $F_D$, a larger length of coherence,
and a smaller vertical phase velocity.

Following Yakovets et al. (1999), we chose to utilize the term
"wave packets" by expanding it to the term "traveling wave
packets" (TWPs). The investigation made in this paper has brought
out clearly that this designation describes most adequately the
phenomenon involved.

Some authors associate the variations of the frequency Doppler
shift $F_D$ with MS TIDs that are generated during the passage of
atmospheric fronts, tornadoes, and hurricanes (Baker and Davies,
1969; Bertin et al., 1975; 1978; Hung et al., 1978; Kersly and
Rees, 1982; Stobie et al., 1983; Huang et al., 1985). It is only
in some cases that these experiments observed quasi-monochromatic
variations of $F_D$ with periods of about 10 min (Huang et al.,
1985).

Thus, in spite of the many years of experimental and theoretical
studied, so far there is no clear understanding not only of the
physical origin of the quasi-monochromatic MS TIDs but even of
their morphology as well (the occurrence frequency as a function
of geographical location, time, level of geomagnetic and
meteorological activity, etc.).

To address these issues requires obtaining statistically
significant sets of experimental data with good spatial
resolution in order to study not only the morphological but also
dynamic characteristics of quasi-monochromatic MS TIDs (the
direction of their displacement, their propagation velocity, and
the location of the possible disturbance source). Another
important requirement implies the continuous, global character of
observations, because such phenomena are temporally highly rare
and spatially random.

Such a possibility is, for the first time, afforded by the use of
the international ground-based network of two-frequency receivers
of the navigation GPS system which at the beginning of 2002
consisted of no less than 1000 sites, with its data posted on the
Internet, which opens up a new era of a global, continuous and
fully computerized monitoring of ionospheric disturbances of a
different class. Analysis and identification of TWPs became
possible through the use of the technology (developed at the
ISTP) for global detection and determination of parameters of
ionospheric disturbances of a different class.

The objective of this paper is to study the morphology and
spatial-temporal properties of TWPs using the data from the global
network of GPS receivers. Section 2 provides general information
about the experiment and gives a brief description of the method
of TWPs detecting. Section 3 presents our new evidence
characterizing the TWPs morphology. Section 4 is devoted
to a detailed analysis of the spatial-temporal properties of TWPs
by considering an example of the most pronounced manifestation of
TWPs on October 18, 2001 as observed in California, USA.
The discussion of our results compared with the findings reported
by other authors is given in Section 5.

\section{General information about the experiment and method of
TWPs detecting}
\label{SPE-sect-2}

This paper presents, for the first time, the morphology of TWPs
for 105 days of 1998--2001, with a different level of geomagnetic
activity and with the number of stations of the global GPS
network ranging from 10 to 300. A total number of he TEC series
used in the analysis, corresponding to the observation along a
single receiver-satellite Line-of-Sight (LOS), with a duration
of each series of about 2.3 hours, exceeded 700000.

For a diversity of reasons, slightly differing sets of GPS
stations were selected for different events to be analyzed;
however, the geometry of experiment for all events was virtually
identical. The stations coordinates are not given here for
reasons of space. This information may be found at the electronic
address http://lox.ucsd.edu/cgi-bin/allCoords.cgi?.
The global GPS covers rather densely North America and
Europe, and to a much lesser extent Asia. GPS stations are more
sparsely distributed on the Pacific and Atlantic Oceans. However,
such coverage of the surface of the globe makes it possible,
already today, to tackle the problem of global detection of
disturbances with hither to unprecedented spatial accumulation.
Thus, in the Western hemisphere the corresponding number of
stations can reach no less than 500 already today, and the number
of LOS to the satellite can be no less than 2000--3000.

The area of California, USA, is particular convenient for our
investigations because of the large number of GPS stations (no
less than 300) located over a relatively small area, which makes
it possible to obtain a great variety of GPS arrays of a
different configuration for a reliable determination of the
dynamic TWPs parameters using the method of GPS interferometry of
TIDs (Afraimovich et al., 1998; 2000).

The comparison of TWPs characteristics with geomagnetic field
variations was based on using the data from the INTERMAGNET
network (INTERMAGNET, http://www.intermagnet.org/).

\subsection{Method of processing the data from the global network.
Selection of TWPs}
\label{SPE-sect-2.1}

The standard GPS technology provides a means for wave
disturbances detecion based on phase measurements of TEC $I_0$
(Hofmann-Wellenhof et al., 1992):

\begin{equation} \label{TWP-eq1}
I_0=\frac{1}{40{.}308}\frac{f^2_1f^2_2}{f^2_1-f^2_2}
                           [(L_1\lambda_1-L_2\lambda_2)+const+nL]
\end{equation}

where $L_1\lambda_1$ and $L_2\lambda_2$ are additional paths of
the radio signal caused by the phase delay in the ionosphere,
(m); $L_1$ and $L_2$ represent the number of phase rotations at
the frequencies $f_1$ and $f_2$; $\lambda_1$ and $\lambda_2$
stand for the corresponding wavelengths, (m); $const$~ is the
unknown initial phase ambiguity, (m); and $nL$~ are errors in
determining the phase path, (m).

Phase measurements in the GPS can be made with a high degree of
accuracy corresponding to the error of TEC determination of at
least $10^{14}$ m${}^{-2}$ when averaged on a 30-second time
interval, with some uncertainty of the initial value of TEC,
however (Hofmann-Wellenhof et al., 1992). This makes possible
detecting ionization irregularities and wave processes in the
ionosphere over a wide range of amplitudes (up to $10^{-4}$ of
the diurnal TEC variation) and periods (from 24 hours to 5 min).
The unit of TEC, which is equal to $10^{16}$ m${}^{-2}$~($TECU$)
and is commonly accepted in the literature, will be used in the
following.

In some instances a convenient way for comparing TEC response
characteristics from the GPS data with those obtained by
analyzing the frequency Doppler shift in the HF range (Davies
and Jones, 1971; Waldock and Jones, 1987; Jacobson and Carlos,
1991; Yakovets et al., 1999) is to estimate the frequency Doppler
shift $F_D$ from TEC series obtained by formula (1).
To an approximation sufficient for the purpose of our
investigation, a corresponding relationship was obtained by
(Davies, 1969):

\begin{equation} \label{TWP-eq2}
F_D=13{.}5\times 10^{-8}I'_t/f
\end{equation}

where $I'_t$ stands for the time derivative of TEC.

Primary data include series of "oblique" values of TEC
$I_0(t)$, as well as the corresponding series of elevations
$\theta(t)$ and azimuths $\alpha(t)$ along LOS to the
satellite calculated using our developed CONVTEC program
which converts the GPS system standard RINEX-files on the
INTERNET (Gurtner, 1993). For TWPs characteristics to be
determined continuous series of $I_0(t)$ series of a duration
of no less than 2{.}3 hours are chosen.

To normalize the response amplitude we converted the "oblique"
TEC to an equivalent "vertical" value (Klobuchar, 1986):

\begin{equation}
\label{TWP-eq3} I = I_0 \times {\rm cos} \left[{\rm
arcsin}\left(\frac{R_z}{R_z + h_{{\rm max}}}{\rm cos}
\hspace{0.1cm} \theta\right) \right],
\end{equation}

where $R_z$ is the Earth's radius, and $h_{max}=300$ km is the height
of the $F_2$-layer maximum.

The most reliable results from the determination of TWPs parameters
correspond to high values of elevations $\theta(t)$ of the beam
to the satellite because sphericity effects become reasonably small.
All results in this study were obtained for elevations $\theta(t)$
larger than 30$^\circ$.

To exclude the variations of the regular ionosphere, as well as
trends introduced by the motion of the satellite, we employ the
procedure of removing the linear trend by preliminarily smoothing
the initial series with a convenient time window.

The technology for global detection of TWPs that was
developed at the ISTP SB RAS makes it possible to select -- from a
large amount of experimental material in the automatic mode -- the
TEC disturbances which can be assigned to a class of TWPs.

The selection of TEC series which could be ascribed to a class of
TWPs was carried out by two criteria (Fig. 1). First of all, TEC
variations were selected, for which the value of the standard
deviation exceeded a given threshold $\epsilon$ (in the present case
$\epsilon$ = 0.1 $TECU$).

In addition, for each filtered series, we verified the fulfillment
of the "quasi-monochromaticity" condition of TEC oscillations, for
which the ratio $R$ of a total spectral signal power in the
selected frequency band $\delta F$ in the neighborhood of a
maximum value of the power $S_{max}$, to a total spectral signal
power outside the frequency band $\delta F$ under consideration
exceeded a given threshold $R_{min}$ (in the present case
$R_{min}$ = 2).

Fig. 1 illustrates the selection process of the TWPs.
Fig. 1a gives an example of weakly disturbed variations of the
"vertical" TEC $I(t)$ as recorded on July 15, 2001 at station DARW
($131.13^\circ$E; $12.8^\circ$S; satellite number PRN05). Fig. 1b
presents the $dI(t)$-variations that were filtered from the
initial $I(t)$-series. Thin horizontal lines show the specified
threshold $\epsilon$. The standard deviation of the
$dI(t)$-variations is 0.019 $TECU$, that is, does not reach
the specified threshold $\epsilon$ = 0.1 $TECU$.

Fig. 1c illustrates the $S(F)$ spectrum of the series $dI(t)$ from
Fig. 1b. Thin vertical lines show the boundaries of the frequency
range $\delta F$. For this spectrum $R$ = 0.66 is smaller than the
specified $R_{min}$ = 2 and, hence, the series $dI(t)$ does not
satisfy the condition of "quasi-monochromaticity".

Fig. 1d, 1e and 1f plots the same dependencies as in Fig. 1a, 1b
and 1c but for station TOW2 ($147^\circ$E; $19.3^\circ$S; satellite
number PRN09). It is evident from Fig. 1d that at the background of
the slow TEC variations there are clearly identifiable (unusual for
background TEC disturbances) oscillations in the form of a wave
packet of a duration of about 1 hour and with a typical period $T$
in the range from 10 to 18 min. The oscillation amplitude of the
detected wave packet exceeds one order of magnitude (as a
minimum) the intensity of background TEC fluctuations of this
range of periods (Afraimovich et al., 2001a).

The relative amplitude of such a response $\Delta I/I_0$ is
considerable, 4 $\%$. As the background value of $I_0$ we used
the absolute "vertical" TEC value of $I_0(t)$ for the site
located at $19{.}3^\circ$S; $147^\circ$E, obtained from IONEX
TEC maps (Mannucci, 1998). Since the main contribution to the
modulation of the TEC is made by the region near the ionospheric
$F$-region peak, the relative amplitude of the local electron
density disturbance $\Delta N/N_0$ in this region can be
several times larger than $\Delta I/I_0$.

It is worthwhile to note that the two examples described above
refer to the same time interval and to the stations spaced by a
distance of no more than 1900 km from one another. This suggests
a local character of the phenomenon and is in agreement with the
overall sample statistic characterizing its spatial correlation
(see below).

The standard deviation of the series $dI(t)$, shown in Fig. 1e,
is 0.114 $TECU$, which is larger than the specified threshold
$\epsilon$=0.1 $TECU$, and this series satisfies the condition
for the standard deviation. Fig. 1f presents the spectrum $S(F)$
of the series $dI(t)$, shown in Fig. 1e. For this spectrum
$R$ = 3.71, which is larger than the specified $R_{min}$ = 2,
that is, in this case the series $dI(t)$ satisfies the condition
of "quasi-monochromaticity".

Panel {\bf e} shows the maximum value of the amplitude
$A_{max}$ of the packet and the time $t_{max}$ corresponding
to this amplitude.

When the filtered dI(t)-series satisfied the conditions described
above, such an event was recognized as TWP.

Furthermore, for each such event, a special file stored
information about the name, geographical latitude $\phi_s$ and
longitude $\lambda_s$ of the GPS station; the GPS satellite PRN
number; the time $t_{max}$ corresponding to the maximum value of
the amplitude $A_{max}$ of the TWP; the amplitude $A_{max}$;
the TWP oscillation period $T$; the $R$ ratio; and about the
value of the elevation $\theta_s(t)$ and the azimuth $\alpha_s(t)$
of the LOS to the satellite calculated for the time $t_{max}$.
The sample statistic, presented below, was obtained by processing
such files for our selected value of $\epsilon$=0.1 $TECU$.

\section{Morphology of TWPs}
\label{SPE-sect-3}

The method outlined above was used to obtain a series of TWPs
totaling about 1300 cases, or about 0.2$\%$ of the total
number of the radio paths considered (receiver-satellite
LOS). As has been pointed out above, the radio paths that were
considered totaled over 700000. An analysis of the resulting
statistic revealed a number of dependencies of the TWPs
parameters on different factors.

First of all, we consider the seasonal dependence of the density
and amplitude of the TWPs (Fig. 2). Fig. 2a plots the
dependence of the number of days of observation $M$ on the time of
the year. It is evident that statistically, the autumn season is
represented best. Fig. 2b shows the seasonal dependence of the
number of TWPs $N$. Fig. 2c plots the number of TWPs $L
= N/M$ per day as a function of time of the year. This dependence
has maxima in winter and in autumn.

The relative TWPs density $D$, obtained as the ratio of the
number of TWPs $N$ to the number of receiver-satellite LOS,
is presented in Fig. 2d. As is apparent from the figure, TWPs
are observed in no more than 0.1-0.4$\%$ of the total number of
radio paths, and much more frequently in winter (over 0.4$\%$) and
autumn (up to 0.3$\%$) than in spring and summer (less than 0.1$\%$).

Diamonds in Fig. 2d show the mean values of $\langle A \rangle$ of
the maximum TWPs amplitudes $A_{max}$ for each season, and vertical
lines show their standard deviations. Thick horizontal line shows
the threshold in amplitude $\epsilon$=0.1 $TECU$. The most probable
value of $\langle A \rangle$ with a small scatter varies around
the value 0.3 $TECU$, irrespective of the season.

Fig. 3d presents the normalized occurrence probability distribution
of TWPs with the specified maximum amplitude of the
packet $A_{max}$. The vertical dashed line shows the threshold
in amplitude $\epsilon$=0.1 $TECU$. It was found that the most
probable value of the amplitude $A_m$, also shown in Fig. 3d, is
about 0.3 $TECU$, and the half-width of the distribution is 0.2
$TECU$. As was shown by Afraimovich et al. (2001a), the mean values
of the TEC variation amplitude with the period of 20 min for the
magnetically quiet and magnetically disturbed days do not exceed
0.01 $TECU$ and 0.07 $TECU$, respectively. Thus the most probable
value of the amplitude $A_m$ of TWPs exceeds the mean
value of the TEC variation amplitude by a factor of 4--6 as a
minimum. This estimate is consistent with the variation amplitude
of the frequency Doppler shift $F_D$ reported by Yakovets et al.
(1999).

Fig. 3b presents the diurnal distribution $P(t_{max})$ of the
times $t_{max}$ corresponding to the maximum value of the
amplitude $A_{max}$ of the wave packet of TWPs. It is
evident that most of the TWPs (about 87$\%$) are observed
during the daytime, from about 7:00 to 16:00 of local time, LT.

Fig. 3a plots the dependence $P(|Dst|)$ of the number of TWPs
on the values of the geomagnetic activity index $Dst$
taken by the modulus. There is a general tendency of the number
of TWPs to increase with the decreasing level of geomagnetic
activity. Most (about 92$\%$) of the TWPs occur when the
values of the $Dst$-index are smaller than 100 nT.

The availability of a large number of GPS stations in some
regions of the globe (in California, USA and West Europe, for
example) makes it possible to determine not only the temporal but
also spatial characteristics of TWPs. To estimate the
radius of spatial correlation of events of this type we
calculated the number of cases where the TWPs within a
single 2.3-hour time interval were observed at any two GPS
stations separated by $dR$. Fig. 3c presents the histogram of
values of $P(dR)$ as a function of distance $dR$. It was found that
the localization of the TWPs in space is strongly
pronounced. In 82$\%$ of cases the distance $dR$ does not exceed
500 km.

\section{Traveling wave packets of total electron content pulsations
as deduced from a case study of the October 18, 2001 event}
\label{SPE-sect-4}

Using the method of GPS interferometry of TIDs (Afraimovich et al.,
1998; 2000), we carried out a detailed analysis of the
spatial-temporal properties of TWPs by considering an example
of the most pronounced manifestation of TWPs on October 18, 2001
over California, USA. Numerous traveling ionospheric disturbances
of the type of TWPs were recorded on that day between 15:00 and
18:00 UT using signals from several satellites, at many GPS
stations located in California.

\subsection{Geometry and general information about the
October 18, 2001 experiment}
\label{SPE-sect-4.1}

The area of California within $220-260^\circ$E; $28-42^\circ$N is
convenient for our investigations because of a large number of GPS
stations located there, which makes it possible to obtain a great
variety of GPS arrays of a different configuration for determining
the TID parameters and provides a means of verifying the reliability
of calculated data. It is also important that for the
above-mentioned time interval 15:18 UT and for the chosen
longitude range the local time varied from 08 to 11 LT, which
reduces the level of background TEC fluctuations characteristic
for the night-time.

Fig. 4 illustrates the geometry of the experiment of October 18,
2001. Heavy dots show the GPS stations, and small dots show the
position of subionospheric points for GPS receiver-satellite LOS.
Since each receiver site observes simultaneously several (no less
than four) GPS satellites, the number of radio paths far exceeds
the number of stations, which enhances the capabilities of
analysis. Fig. 4a presents the entire set of GPS stations used in
the experiment. Fig. 4b and 4c show the stations and the
subionospheric points where the TEC variations revealed
TWPs with an amplitude exceeding the specified threshold
$\epsilon$=0.05 $TECU$ (Fig. 4b) and $\epsilon$=0.1 $TECU$ (Fig. 4c).
With the sole exception, TWPs were recorded along the paths over and,
predominantly in the north-eastward direction. As is evident from
Fig. 4, an increase of the recording threshold by a factor of
two reduced the number of recorded events by a factor of two.
Stations shown in Fig. 4c were used as the elements of the GPS arrays in
calculating the TWPs parameters.

The geomagnetic situation on October 18, 2001 may be
characterized as a weakly disturbed one, which must lead to some
increase of the level of background TEC fluctuations yet cannot
cause large-scale changes in electron density characteristic for
a geomagnetically disturbed ionosphere.

Geomagnetic field $Dst$-variations for October 18, 2001 are
plotted in Fig.~5a. In the analysis of the geomagnetic situation we
used also the data from magnetic observatory Victoria ($48.52^\circ$N;
$236.58^\circ$E) where for the time interval of our interest a weak
geomagnetic disturbance was recorded, which implied a decrease of the
horizontal component $H(t)$ of the magnetic field by 60 nT (Fig.
5b). There was a concurrent, small decrease of the fluctuation
amplitude of the $dH(t)$-component in the range of periods 2-20 min
(Fig. 5c). The range of variation of the geomagnetic $Dst$-index
for the selected time interval was also relatively small (no more
than 20 nT), yet the period 15:30-18:00 UT showed a clearly
pronounced decrease of the $Dst$-index coinciding with the period
of the decrease of the $H$-component of the magnetic field (Fig.~5a).

Fig. 5d presents the distribution $N(t)$ of the number of
TWPs detected for that day by all stations of the global
GPS system analyzed here, with the standard deviation in excess
of $\epsilon$ = 0.1 $TECU$. Fig. 5e illustrates the dynamic amplitude
spectrum $S(f,t)$ of TEC variations in the range of periods 5--60
min obtained by using the method of spatial averaging of the
spectra for the entire California region (Afraimovich et al.,
2001a).

Overall, the TEC variations correlate with geomagnetic field
variations. Between 15:00 and 19:00 UT, the enhancement of the
oscillations of the $H$-component was accompanied by an expansion
of the spectrum and by an increase of the TEC fluctuation
amplitude. The highest intensity is shown by the TEC oscillations
with periods of 12--17 min between 15:30 and 17:00 UT. The largest
number of TWPs was also recorded during the same period
of time (Fig.~5d).

To check that TWPs were observed on that day somewhere else on the
globe and not only between 15 and 19 UT, we processed the data
with different values of the threshold $\epsilon$ for the entire
global GPS network.

Fig.~6 presents the sample statistic of the pulsations identified
for October 18, 2001 as a function of UT and local time LT,
calculated for the longitude of $240^\circ$E corresponding to the middle
of the California region: Fig. 6a - from TWPs with the standard deviation
higher than $\epsilon$ = 0.1 $TECU$ obtained from all stations of
the global GPS network used in the study (copy of Fig. 5d); Fig. 6b and
Fig.~6c - same as in Fig.~6a but for $\epsilon$ = 0.05 $TECU$ and
$\epsilon$ = 0.01 $TECU$; Fig.~6d - standard deviation higher than
$\epsilon$ = 0.01 $TECU$ for the data from the California region only.

An analysis of the Fig.~6 data leads us to conclude that the TWPs
on that day were observed mainly in California only and only
over the time interval 15--17 UT.

\subsection{Methods of determining the form and dynamic
characteristics of TWPs}
\label{SPE-sect-4.2}

The methods of determining the form and dynamic characteristics
of TIDs that are used in this study are based on those reported
in (Mercier, 1986; Afraimovich, 1995; 1997; Afraimovich et al.,
1998; 1999).

We determine the velocity and direction of motion of the phase
interference pattern (phase front) in terms of some model of this
pattern, an adequate choice of which is of critical importance.
In the simplest form, space-time variations in phase of the
transionospheric radio signal that are proportional to TEC
variations $I(t, x, y)$ in the ionosphere, at each given time $t$
can be represented in terms of the phase interference pattern that
moves without a change in its shape (the non dispersive
disturbances):

\begin{equation}
I(t,x,y)=F(t-x/u_x-y/u_y)
\label{TWP-eq4}
\end{equation}

where $u_x(t)$ and $u_y(t)$ are the velocities of intersection of the
phase front of the axes x (directed to the East) and y (directed to
the North), respectively.

A special case of (4) is the most often used model for a
solitary, plane travelling wave of TEC disturbance:

\begin{equation}
I(t,x,y)=\delta\sin(\Omega t-K_xx-K_yy+\varphi_0)
\label{TWP-eq5}
\end{equation}

where $I(t,x,y)$ are space-time variations of TEC;
$\delta(t)=\exp[-\left(({t-t_{{\rm max}}})/({t_{{\rm d}}})\right)^2]$
-- the amplitude; $K_x$, $K_y$, $\Omega$ are the
$x$- and $y$- projections of the wave vector {\boldmath $K$}, and
the angular frequency of the disturbance, respectivly;
$\varphi_0$ is the initial disturbance phase;
$t_{{\rm max}}$ is the time when the disturbance has a maximum
amplitude; $t_{{\rm d}}$ is the half-thickness of the 'wave packet'.

It should be noticed, however, that in real situations neither
of these ideal models (4), (5) are realized
in a pure form. This is because that the AGW that cause TIDs
propagate in the atmosphere in the form of a dispersing wave
packet with a finite value of the width of the angular spectrum.
But in the first approximation on short time interval of
averaging compared to time period of filtered variations of TEC,
the phase interference pattern moves without a substantial
change in its shape.

Mercier (1986) suggested a statistical method to analyze the
phase interference pattern. Primary data comprise time
dependencies of the spatial phase derivatives $I'_y(t)$ and
$I'_x(t)$ along the directions $y$ and $x$. Method Mercier
(1986) involves determining a series of instantaneous values
of the direction $\alpha(t)$

\begin{equation}
\alpha(t)=\arctan(I'_x(t)/I'_y(t))
\label{TWP-eq6}
\end{equation}

and constructing subsequently, on a chosen time interval,
the distribution function of azimuth $P(\alpha)$. The central
value of $\alpha$ is used by Mercier as an estimate of the azimuth
of prevailing propagation of TIDs (modulo 180${}^\circ$).

The other method is based on analyzing the phase
interference pattern anisotropy in the antenna array plane
by determining the contrast $C$ of the interference pattern
(Mercier, 1986). In this case the ratio $C_{x,y}$ is
calculated as follows:

\begin{equation}
\begin{array}{rlrl}
C_{x,y}&=\sigma_X/\sigma_Y, & \mbox{if } \sigma_X&>\sigma_Y\\
C_{x,y}&=\sigma_Y/\sigma_X, & \mbox{if } \sigma_Y&>\sigma_X
\end{array}
\label{TWP-eq7}
\end{equation}

where $X$ and $Y$ are series of the transformed values of
$I'_x(t)$ and $I'_y(t)$ obtained by rotating the original
coordinate system ($x$, $y$) by the angle $\beta$:

\begin{equation}
\begin{array}{rl}
X(t)&=I'_x(t)\sin\beta+I'_y(t)\cos\beta\\
Y(t)&=-I'_x(t)\cos\beta+I'_y(t)\sin\beta \end{array}
\label{TWP-eq8}
\end{equation}

and $\sigma_X$ and $\sigma_Y$ are r. m. s. of the corresponding
series.

Mercier (1986) showed that it is possible to find such a value
of the rotation angle $\beta_0$, at which the ratio $C_{x,y}$
will be a maximum and equal to the value of contrast $C$. This
parameter characterizes the degree of anisotropy of the phase
interference pattern. The angle $\beta_0$ in this case
indicates the direction of elongation, and the angle
$\alpha_c=\beta_0+\pi/2$ indicates the direction of the wave
vector {\boldmath $K$} coincident (modulo 180${}^\circ$) with
the propagation direction of the phase front.

The method of Mercier (1986) essentially makes it possible
to determine only the anisotropy and the direction of
irregularity elongation of the phase interference pattern
(modulo 180${}^\circ$).

A Statistical, Angle-of-arrival and Doppler Method (SADM)
is proposed by Afraimovich (1995, 1997) for determining
characteristics of the dynamics of the phase interference
pattern in the horizontal plane by measuring variations of
phase derivatives with respect not only to spatial
coordinates $I'_x(t)$ and $I'_y(t)$ - proportional to
the angles of arrival variations, but additionally to
time $I'_t(t)$ - proportional to the frequency Doppler
shift variations. This permits to ascertain the unambiguous
orientation of $\alpha(t)$ of the wave-vector {\boldmath $K$}
in the range 0--360${}^\circ$ and determine the horizontal velocity
modulus $V_h(t)$ at each specific in-stant of time.

Afraimovich et al.(1998, 1999) described updating of the SADM
algorithm for GPS-arrays (SADM-GPS) based on a simple model for
the displacement of the phase interference pattern that travels
without a change in the shape and on using current
information about angular coordinates of the GPS satellites:
the elevation $\theta_s(t)$ and the azimuth $\alpha_s(t)$.
Of course, such an approximation is acceptable only for large
values of the LOS elevation $\theta_s$.

The method SADM-GPS permits to determine the horizontal velocity
$V_h(t)$ and the azimuth $\alpha(t)$ of TID displacement
at each specific in-stant of time (the wave-vector orientation
{\boldmath $K$}) in a fixed coordinate system ($x$, $y$):

\begin{equation}
\begin{array}{rl}
\alpha(t) &=\arctan(u_y(t)/u_x(t))\\
u_x(t) &=I'_t(t)/I'_x(t) = u(t)/\cos\alpha(t) \\
u_y(t) &=I'_t(t)/I'_y(t) = u(t)/\sin\alpha(t) \\
u(t) &=|u_x(t) u_y(t)|/(u_x^2(t) + u_y^2(t))^{-1/2} \\
V_x(t) &=u(t)\sin\alpha(t)+w_x(t) \\
V_y(t) &=u(t)\cos\alpha(t)+w_y(t) \\
V_h(t) &= (V_x^2(t)+V_y^2(t))^{1/2}
\end{array}
\label{TWP-eq9}
\end{equation}

where $u_x$ and $u_y$ are the propagation velocities of the phase
front along the axes $x$ and $y$ in a frame of reference related to
the GPS-array; $w_x$ and $w_y$ are the $x$ and $y$ projections of the
velocity $w$ of the subionospheric point (for taking into
account the motion of the satellite).

The coordinates of the subionospheric point $x_s(t)$ and $y_s(t)$
at $h_{max}$ in the chosen topocentric coordinate system vary as:

\begin{equation}
\begin{array}{rl}
x_s(t) = h_{max} sin(\alpha_s(t)) ctg(\theta_s(t)) \\
y_s(t) = h_{max} cos(\alpha_s(t)) ctg(\theta_s(t))
\end{array}
\label{TWP-eq10}
\end{equation}

and the $x$- and $y$- components of the displacement velocity $w$:

\begin{equation}
\begin{array}{rl}
w_x(t)=d/dt(x_s(t)) \\
w_y(t)=d/dt(y_s(t)) \\
w(t)=(w_x^2(t)+ w_y^2(t))^{1/2}
\end{array}
\label{TWP-eq11}
\end{equation}

Let us take a brief look at the sequence of data handling
procedures. Out of a large number of GPS stations, three
points (A, B, C) are chosen in such a way that the
distances between them do not exceed about one-half the expected
wavelength $\Lambda$ of the disturbance. The point B is
taken to be the center of a topocentric frame
of reference. Such a configuration of the GPS receivers
represents a GPS-array (or a GPS-interferometer) with a minimum
of the necessary number of elements. In regions with a dense
network of GPS-points, we can obtain a broad range of GPS-arrays
of a different configuration, which furnishing a means of testing
the data obtained for reliability; in this paper we have taken
advantage of this possibility.

The input data include series of the vertical TEC $I_A(t)$,
$I_B(t)$, $I_C(t)$, as well as corresponding series of
values of the elevation $\theta_s(t)$ and the azimuth
$\alpha_s(t)$ of the LOS.

Series of the values of the elevation $\theta_s(t)$ and azimuth
$\alpha_s(t)$ of the LOS are used to determine the location of
the subionospheric point, as well as to calculate the elevation
$\theta$ of the wave vector $\boldmath K$ of the disturbance
from the known azimuth $\alpha$ (see formula (12)).

Since the distance between GPS-array elements (from several tens
to a few hundred of kilometers) is much smaller than that to
the GPS satellite (over 20000~km), the array geometry at the height
near the main maximum of the $F_2$-layer is identical to that on the
ground.

Linear transformations of the differences of the values of the
filtered TEC $(I_{{\rm B}}-I_{{\rm A}})$ and
$(I_{{\rm B}}-I_{{\rm C}})$ at the receiving
points A, B and C are used to calculate the components of the
TEC gradient $I'_x$ and $I'_y$ (Afraimovich et al., 1998).
The time derivative of TEC $I'_t$ is determined by differentiating
$I_{{\rm B}}(t)$ at the point B.

The resulting series are used to calculate instantaneous
values of the horizontal velocity $V_h(t)$ and the azimuth
$\alpha(t)$ of TID propagation. Next, the series $V_h(t)$ and
$\alpha(t)$ are put to a statistical treatment. This involves
constructing distributions of the horizontal velocity $P(V_h)$
and direction $P(\alpha)$ which are analyzed to test the hypothesis
of the existence of the preferred propagation direction. If such a
direction does exist, then the corresponding distributions are
used to calculate the mean value of the horizontal velocity
$\langle V_h\rangle$, as well as the mean value of the
azimuth $\langle \alpha\rangle$ of TID propagation.

As a first approximation, the transionospheric sounding
method is responsive only to TIDs with the wave vector
$\boldmath K$ perpendicular to the direction $\boldmath r$
of the LOS. A corresponding condition for elevation $\theta$
and azimuth $\alpha$ has the form

\begin{equation}
\tan\theta=-\cos(\alpha_s-\alpha)/\tan\theta_s
\label{TWP-eq12}
\end{equation}

Hence the phase velocity modulus $V$ can be defined as

\begin{equation}
\label{TWP-eq13}
V=V_h\times cos(\theta)
\end{equation}

The aspect dependence (12) of the TEC disturbance amplitude
is of significant importance for investigating wave disturbances. The
condition (12) imposes a constraint on the number of
LOS for which a reliable detection of TIDs at the background of noise is
possible. The aspect effect causes the disturbance maximum to be
displaced along the time axis, which can introduce errors in
determining the TID displacement if the velocity is calculated
from time delays. Furthermore, as will be shown below, the aspect
effect will give rise to structures of the type of wave packets
"observed" in TEC variations, which do not exist in the reality.

Afraimovich et al. (1992) showed that for the Gaussian ionization
distribution the TEC disturbance amplitude $M$ is determined by
the aspect angle $\gamma$ between the vectors $\boldmath K$ and
$\boldmath r$, as well as by the ratio of the wavelength of the
disturbance $\Lambda$ to the half-thickness of the ionization
maximum $h_d$:

\begin{equation}
M(\gamma) \propto \frac {h_d}{sin(\theta_s)} \cdot
\exp \left[ -\left( \frac {\pi h_d \cos(\gamma)}
{\Lambda \sin(\theta_s)} \right)^2 \right]
\label{TWP-eq14}
\end{equation}

In this paper the influence of aspect effects on the character of
TEC behavior and on the accuracy of the calculated parameters of
TWPs was investigated and the reliability of the
determination of TWPs characteristics was verified by modeling the
wave disturbances of electron density for the observing
conditions of October 18, 2001.

Thus, on the basis of using the transformations described in this
section, for each of the GPS arrays chosen for the analysis we
obtained the average (for the time interval of about 1-2 hours)
values of the following TWPs parameters: $\langle A_I\rangle$, the
amplitude of the TEC disturbance; $\langle \alpha\rangle$ and $\langle
\theta\rangle$ -- the azimuth and elevation of the wave vector
$\boldmath K$; $\langle V_h\rangle$ and $\langle V\rangle$ --
the horizontal component and the phase velocity modulus,
the contrast $C$, and the azimuth of a normal to the wave front
$\alpha_c$ from the method reported by Mercier (1986).

\subsection{Dynamic characteristics of TWP}
\label{SPE-sect-4.3}

Fig. 7a plots the typical time dependencies of TEC $I(t)$ for GPS
satellite PRN14 for three GPS stations: BRAN, CHMS, and DUPS in
California. The three GPS sites constitute a typical GPS array,
the data from which were processed by the technique described in
the preceding section. For the same stations Fig. 7b presents the
TEC variations $dI(t)$ that were filtered from the initial series
$I(t)$ using the band-pass filter with the boundaries from 5 to 20
min.

The filtered series for the period 15:00--16:00~UT show the
presence of significant TEC pulsations of the type of single wave
packet with a duration of about one hour and with the amplitude $A$
= 0.5~$TECU$. Not only does the range of the filtered oscillations
$dI(t)$ far exceed the error of phase measurements
($10^{-3}$~$TECU$), but it also exceeds nearly an order of magnitude
the level of background TEC variations. TEC variations from the
three spatially separated GPS stations show a high degree of
similarity and have a small time shift. This suggests that we
are dealing here with the same traveling disturbance.

Results of a calculation (using the SADM-GPS algorithm) of the
mean values of the azimuth $\langle \alpha\rangle$ and the
horizontal velocity $\langle V_h\rangle$ of the disturbance for
each 30-s time interval are presented in Fig. 7c and 7d. As is
evident from the figure, this wave packet was traveling
predominantly in the south-eastward direction with the mean
velocity of about 180 m/s. The scatter of the counts is caused
by the incomplete correspondence of the actual picture of an ideal
TWP model in the form of a monochromatic packet (5)
and, in particular, by the presence of background non-correlated
TEC fluctuations (Afraimovich et al., 1998).

A processing of the data from the other GPS arrays in the same
region (Fig.~4c) by use of the SADM-GPS method provided
distributions of the main TWP parameters recorded on October 18,
2001 in California. The various combinations of GPS arrays for
the time interval 15--16 UT totaled 231. Statistical data show an
agreement of the mean values of the calculated parameters to
within their standard deviations, which indicates a good
stability of the data obtained, irrespective of the particular
configuration of a GPS array.

Distributions of the mean values of the TWP parameters calculated
for each of the 231 GPS arrays are presented in panels ({\bf a}~--
{\bf c}) of Fig. 8. According to our data, the value of the horizontal
propagation velocity $V_h$ of TWP (Fig. 8a) varies from 60 to 270~m/s,
on the average, with the most probable value 190~m/s. The
TWP wavelength $\Lambda$ with the mean oscillation period of about
1000 s is on the order of 150--200~km.

An analysis of the distribution of the azimuths $P(\alpha)$
(Fig.~8b) shows a clearly pronounced south-eastward direction of TWP
displacement $140\pm20^\circ$. The average (for 231 GPS arrays) value
of the contrast $C$ = 7 suggests a strong anisotropy of TWP. Fig. 8b
shows also the distribution of the azimuth of a normal to the
wave front $P(\alpha_c)$ deduced by the method of Mercier (1986).
This distribution virtually coincides with $P(\alpha)$, suggesting
that the TWP travel across their elongation. Thus the typical size
of the entire wave packet along the propagation direction is about
300--500~km, and along the wave front it is as long as 1000 m.

The arrow in Fig. 4c schematically shows the wave vector $\boldmath K$
of the TWP. The values of $\alpha$ and $V_h$, presented in Fig. 4c,
correspond to the most probable values of the propagation azimuth and
the modulus of the horizontal velocity of the TWP.

The elevation of the TWP wave vector, determined from the aspect
condition (12), has mostly a small positive value
(Fig.~8c): i.e. the oscillation phase propagates upward. Since the
mean value of the elevation is only $+10^\circ$, it can be assumed
that TEC pulsations represent an almost horizontal wave. Accordingly,
estimations of the modulus of the phase velocity $V$ (Fig.~8a) give
values close to the value of its horizontal projection 50--270~m/s,
with the largest probable value 180~m/s.

\subsection{Modeling}
\label{SPE-sect-4.4}

In this paper the influence of aspect effects on the character of
TEC behavior and on the accuracy of the calculated of TEC
pulsations was investigated and the reliability of the
determination of the TWP characteristics was verified by modeling
the wave disturbances of electron density for the observing
conditions of October 18, 2001.

Our developed model of TEC measurements with the GPS
interferometer makes it possible to calculate as realistic a
spatial and temporal distribution of the local electron density
$N_e$ in the ionosphere as possible and then, using the coordinates
of the receiver sites and of the satellites, to integrate $N_e$
along the receiver-satellite LOS with a given step over time
(Afraimovich et al., 1998). As a result we obtain time series of
TEC similar to input experimental data which can be processed by
the same methods as used to process experimental data.

The ionization model takes into account the height distribution
of $N_e$, the seasonal and diurnal density variations determined by
the zenith angle of the Sun, as well as irregular disturbances of
$N_e$ of a smaller amplitude and smaller spatial scales in the form
of a discrete superposition of plane traveling waves.

In this paper, with the purpose of elucidating the origin of TWP,
three types of disturbances were modeled:

a) the disturbance in the form of a single plane wave with amplitude
$A_1$=3\% in the ionization maximum, with the period $T_1$=15~min and
the wavelength $L_1$=172~km. The elevation of the wave
$\theta_1$=10${}^\circ$, and the azimuth $\alpha_1$=146${}^\circ$;

b) the disturbance in the form of a superposition of two plane
waves with periods $T_1$=15~min and $T_2$=12~min, and with the
wavelengths $L_1$=172~km and $L_2$=138~km. The elevations, azimuths
and azimuths of the waves were specified identical:
$\theta_1$=$\theta_2$=10${}^\circ$, $\alpha_1$=$\alpha_2$=146${}^\circ$,
$A_1$=$A_2$=3\% of the value of $N_e$ in the ionization maximum;

c) the disturbance in the form of a single wave packet, with the
semi-thickness $t_d$=20~min and a maximum amplitude at the time
$t_{max}$=15.5~UT. The oscillations inside the packet had the period
$T_1$=15~min and the wavelength $L_1$=172~km.

Disturbance parameters were taken to be close to those obtained
from experimental data using the technique from (Afraimovich et
al., 1998). TEC pulsations were modeled with the purpose of (1)
verifying the reliability of the calculated TWP characteristics,
and (2) elucidating the origin of TWP. A detailed description of
the model used is given in (Afraimovich et al., 1998).

Fig. 7 (panels {\bf e}--{\bf h} at the right) presents the results
of calculations in terms of the model of the TWP in the form of a
single wave packet for the BRAN, CHSM, DUPR array on October 18,
2001. Parameters of the wave packet were taken to be close to
experimental data (Fig.~7a--d). Distributions of pulsation
parameters obtained in a similar modeling of TWP for the other
California GPS arrays are presented in Fig. 8 (panels
{\bf d}--{\bf f} at the right).

Fig. 7 illustrates a good similarity of the experimental and
model TEC variations $I(t)$, $dI(t)$ and the dependencies $V_h(t)$
and $\alpha(t)$. Noteworthy is the weaker inclination of the model
TEC $I(t)$ (Fig. 7e) when compared with the experimental one (Fig. 7a).
This is because for the sake of simplicity and for illustrative
purposes, diurnal variations in ionization in the model are
proportional to the cosine of the zenith angle of the Sun. In
actual conditions the dependence is more complicated, which gives
a faster temporal growth of the TEC. because the trend is removed
in the subsequent discussion and only the relative TEC variations
$dI(t)$ are considered, the above-mentioned difference in the
behavior of the model and experimental TEC will not affect
results obtained.

A comparison of the TEC disturbance parameters specified in the
model with the corresponding values obtained following a
processing by formulas of the SADM-GPS method shows a relatively
good agreement of these values. The azimuth of the wave vector of
TWP was taken to be $\alpha=146^\circ$, and the mean value of the
azimuth calculated by the SADM-GPS method is $146.2^\circ$ (Fig.~7g,
and Fig.~8e); the horizontal velocity was specified equal to 180~m/s,
the calculated mean value of the velocity was 207.5~m/s (Fig. 7h, and
Fig.~8d); the elevation in the model was $\theta=10^\circ$, and the
most probable value of the calculated elevation was $10^\circ$
(Fig.~8f). The azimuth of TWP propagation determined by the SADM-GPS
method was also close to the azimuth values calculated by analyzing
the contrast of the phase interference pattern (Mercier, 1986) --
Fig.~8b,~e. All this demonstrates the validity of the SADM-GPS
technique and confirms the reliability of the TWP parameters
obtained using this technique.

Let us now consider the possible mechanisms that are responsible
for the formation of structures of the type of pulsations in
observed TEC variations. The recorded TEC pulsation represents a
single wave packet with a duration of about 60 min, the
oscillation period inside the packet of 12--15 min, and with the
amplitude exceeding the level of background fluctuations by a
factor of 2--5 (Fig.~7b). Structures of such a type can be
produced in TEC variations in different ways.

If a monochromatic wave with a period of about 15 min (the period
of observed TEC variations) propagates in the ionosphere, then
TEC oscillations of the type of pulsations could be produced
through the aspect effect. Fig. 9a, d illustrates such a
situation. Panels {\bf a} and {\bf d} present the filtered TEC series
obtained by modeling a single plane wave under different
conditions of its observation. At the BRAN site (Fig. 9a) the
elevations of the PRN14 satellite are close to 90$^\circ$, and the
wave vector of the disturbance is perpendicular to the
receiver-satellite LOS throughout the observing period. This is
indicated by the character of the theoretical dependence $M(\gamma)$
calculated by formula (14) for this LOS: $M(\gamma)$
is close to 1 over the entire observing interval (Fig.~9c).
Waves disturbances of the TEC close to the specified monochromatic wave
are therefore observed along the BRAN--PRN14 LOS.

At the BRU1 site for PRN05, the detection conditions for
disturbances are significantly worse. In this case for the first
half an hour the wave propagates virtually along the LOS
($M(\gamma)<0.3$, Fig. 9f), and at the end of the observing interval he
BRU1--PRN15 LOS is perpendicular to the propagation direction of
the wave but has low elevations ($M(\gamma)>1$, Fig. 9f). Owing to
this, TEC variations show an increase in the oscillation
amplitude from 0 to 0.7 $TECU$ -- there arises a feature resembling
a wave packet that was absent in the initial specified model of
the wave. However, as is evident from the figure, in this case
the increase in the TEC variation amplitude has an almost linear
character while the experiment (Fig.~7b) shows a nonlinear
amplitude modulation.

An investigation of the character of the aspect dependence for
all GPS arrays that were used in this study, showed that in 97\%
of cases (including those shown in Fig. 7b) the situation is
realized, which is depicted in Fig. 9c, i.e. $M(\gamma)$ is close to 1
throughout this observing interval. Considering all that has been
said above, one can draw the conclusion that the recorded TEC
pulsations are indeed caused by the propagation of an actual wave
packet of a local density disturbance in the ionosphere rather
than resulting from the recording conditions.

Another possibility of the production of a TWP is the
combination of two or several monochromatic waves with close
periods propagating in the ionosphere (Yakovets et al., 1999). We
modeled the propagation of two quasi-horizontal waves with
periods $T_1$=15~min and $T_2$=12~min by identical amplitudes,
velocities and directions of propagation. The resulting TEC
variations along two LOS: BRAN--PRN14, and BRU1--PRN05, with the
trend removed, are presented in Fig. 9b and 9e, respectively.

As would be expected, there arise amplitude-modulated TEC
oscillations with the modulation period of about
$T=\frac{T_1 T_2}{T_2-T_1}$=60~min. The period is close to
the length of the wave packet recorded experimentally; however,
a simple superposition of two waves gives not one but
a whole chain of wave packets. In the case where the aspect
effect affects little the character of TEC variations (Fig.~9c)
these packets have, in addition, different amplitudes as well
(Fig.~9b). A phase change of the combined waves affects little
the picture presented here by altering the form and position
of the amplitude minimum only slightly. The aspect dependence
in Fig.~9f, however, introduces an additional modulation into
the recorded signal - there occurs a significant enhancement
of one of the chain's packets (Fig.~9e), and the picture
approaches what is observed experimentally in Fig.~7b.

Nevertheless, it cannot be believed that the observed single TWPs
are the result of the aspect effect. Firstly, in the
case of the aspect modulation, because of the slow(nearly linear)
change of the amplitude, not one but at least two wave packets
are observed. Secondly, as has been pointed out above, the
character of the aspect dependence in most cases of TWP
recordings was such that it affected little the amplitude of
disturbances. To obtain the closets TEC oscillations to those
observed experimentally we had to introduce an artificial
modulation, i.e. the amplitude of the initial monochromatic wave
of the disturbance $N_e$ was specified not constant but it had a
time dependence in the form of a Gaussian function. It is such a
model of TEC variations that is shown in Fig.~7f.

\section{Discussion of results}
\label{SPE-sect-5}

Sequences of wave packets similar to the ones shown in Fig.~6 at
the left, were recorded in frequency Doppler variations with the
multifrequency Doppler ionosonde at Almaty (Yakovets et al.,
1999). The recordings of the frequency Doppler shift presented
clearly show chains of wave packets (on the average, 2--3 wave
packets with the record length of about 8~hours). The duration of
the wave packets averaged 90~min, and the oscillation period in a
packet averaged 16~min.

The authors of the cited reference observed two types of
FD-variations: quasi-stochastic TIDs, and monochromatic TIDs in
the form of wave packets. They concluded that quasi-stochastic
TIDs are characterized by a random behavior of the phase, a small
length of coherence, and by a large vertical phase velocity. Wave
packets show quasi-monochromatic oscillations, a larger length of
coherence, and a smaller vertical phase velocity.

Our estimate of the radius of spatial correlation (on the order
of several hundred kilometers) is in reasonably good agreement
with the data reported by Yakovets et al. (1995).

Yakovets et al. (1999) argued that the observed wave packets are
the superposition of the direct and ground-reflected wave whose
source lies in the troposphere. The analysis made in this paper
did not confirm the validity of such an explanation for the
October 18, 2001 TWP.

By comparing our detected TEC pulsations with the data from
(Yakovets et al., 1999; Hines, 1960; Waldock and Jones, 1984;
Francis, 1974) as well as with the findings of our modeling, it
can be assumed that in the atmosphere there exists an additional
amplitude modulation mechanism for wave processes which make sit
possible to obtain TEC oscillations in the form of a single wave
packet. Francis (1974), by considering the auroral electrojet to
be the source of TIDs, showed that upon propagating through the
atmosphere into the $F$-region, the ground-reflected waves acquire
the properties of a wave packet. On the basis of the dispersion
relation Hines (1960) and Waldock and Jones (1984) showed that
TIDs that are associated with a tropospheric jet flow occur in
the $F$-region in the form of a wave packet with
quasi-monochromatic oscillations, the period of which is a
function of inclination angle of the wave vector during the
propagation of the wave from the source to the place of
observation in the $F$-region.

Our data on the TWP displacement velocity and direction
correspond to those of mid-latitude medium-scale traveling
ionospheric disturbances (MS TIDs) obtained previously in the
analysis of phase characteristics of HF radio signals (Kalikhman,
1980; Afraimovich, 1982; Waldock and Jones, 1984; 1986; 1987;
Jacobson and Carlos, 1991), as well as signals from
first-generation navigation satellites (Spoelstra, 1992),
geostationary satellites (Afraimovich et al., 1997b; Jacobson et
al., 1995) and discrete space radio sources (Mercier, 1986; 1996;
Velthoven et al., 1990).

\section{Conclusion}
\label{SPE-sect-6}

Main results of this study may be summarized as follows:

\begin{enumerate}

\item The most of the TWPs are observed during the daytime in no
more than 0.1-0.4$\%$ of the total number of radio paths,
most commonly in winter and autumn in a weakly disturbed
or quiet geomagnetic situation. The distance between
any two GPS stations where the TWPs within a single 2.3-hour
time interval were observed does not exceed 500 km.

\item TWPs in the time range represent quasi-periodic oscillations
of TEC of a length on the order of 1 hour with the oscillation
period in the range 10--20 min and the amplitude exceeding the
amplitude of "background" TEC fluctuations by one order of
magnitude, as a minimum and is 0.3 $TECU$.

\item The dynamical parameters of the TWP observed on October 18,
2001 over California, USA were determined. The TWP traveled with the
elevation $\theta$ = 10${}^\circ$ and the azimuth $\alpha$ =
146${}^\circ$. Its mean velocity $\langle V\rangle$ = 180 m/c
corresponds to the velocity of medium-scale AGW.

\item It is probably that TWP origins are medium-scale AGW.

\end{enumerate}

\section*{ACKNOWLEDGMENTS}

We are indebted to Dr A.S. Potapov for participating in discussions.
We thank O.S. Lesyuta for help in organizing the experiment. We are
also grateful to V.G. Mikhalkovsky for his assistance in preparing
the English version of the manuscript. This work was
done with support under RFBR grant of leading scientific schools
of the Russian Federation No. 00-15-98509 and INTAS grant
No. 99-1186 as well as Russian Foundation for Basic Research
grants No. 01-05-65374, 00-05-72026.

\end{document}